\documentclass[pre,twocolumn,a4paper,showkeys,showpacs]{revtex4-1}
\usepackage[dvipdfmx]{graphicx}
\usepackage{bmpsize}
\usepackage{calc}
\usepackage{amsmath}
\usepackage{physics}
\usepackage{bm}         
\usepackage{mathrsfs}  
\usepackage{amssymb}
\usepackage{comment}
\usepackage{cases}
\usepackage{here}
\usepackage{array}
\usepackage[dvipdfmx]{hyperref}
\usepackage{xr}
\usepackage{xcolor}
\usepackage{tabularx}
\newcolumntype{C}{>{\centering\arraybackslash}X}
\newcolumntype{L}{>{\raggedright\arraybackslash}X}
\newcolumntype{R}{>{\raggedleft\arraybackslash}X}
\usepackage{ulem}

\usepackage{lipsum}

\newcommand{\leri}[1]{\left({#1}\right)}
\newcommand{\sgn}[1]{\mathrm{sgn}\leri{{#1}}}

\begin{document}

\title{
Boltzmann approach to collective motion via nonlocal visual interaction
}
\author{Susumu Ito and Nariya Uchida}
\thanks{uchida@cmpt.phys.tohoku.ac.jp}
\affiliation{Department of Physics, Tohoku University, Sendai, 980-8578, Japan}
\date{\today} 

\begin{abstract}
Visual cues play crucial roles in the collective motion of animals, birds, fish, and insects.
The interaction mediated by visual information is essentially non-local and 
has many-body nature due to occlusion, which poses a challenging problem
in modeling the emergent collective behavior.
In this paper, we introduce a Boltzmann-equation approach incorporating 
non-local visual interaction.
Occlusion is treated in a self-consistent manner via a coarse-grained density field, 
which renders the interaction effectively pairwise.
Our model also incorporates the recent finding 
that each organism stochastically selects a neighbor to interact at each instant.
We analytically derive the order-disorder transition point,
and show that the visual screening effect raises the transition threshold,
which does not vanish when the density of the agents or 
the range of the intrinsic interaction is taken to infinity.
Our analysis suggests that the model exhibits 
a discontinuous transition as in the local interaction models, 
and but the discontinuity is weakened by the non-locality.
Our study clarifies the essential role of non-locality in the visual interactions among moving organisms.
\end{abstract}

\maketitle

%%%%%%%%%%%%%%%%%%%%%%%%%%%%%%%%%%%%%%%%%%%%%%%%%%%%%%%%%%%%
\section{Introduction}
Collective motion is ubiquitously found in Nature~\cite{Vicsek2012}.
Visual information plays an important role in the interaction
between organisms that have eyes:
insects~\cite{Sridhar2021}, fish~\cite{Sridhar2021,Peshkin2013}, 
birds~\cite{Pearce2014}, and humans~\cite{Moussaid2011}.
In the visual interaction, each organism often selects
a specific neighbor to interact with and decides its next motion~\cite{Sridhar2021,Moussaid2011,Herbert2011}.
The selective decision-making reduces the load on the information 
processing system in the brain~\cite{Dukas1998}.
A recent study proposes a mechanism of stochastic pairwise interaction, 
in which each individual randomly selects another 
and copy its orientation~\cite{Jhawar2020}. 

Agent-based models are a powerful tool to
simulate collective motion~\cite{Vicsek1995,Couzin2002,chate2008collective}.
Some of them consider visual information
under the assumption that each agent interacts 
with all detected neighbors simultaneously~\cite{Pearce2014,Lemasson2009,Kunz2012,Bastien2020,Castro2024}.
Several other models incorporate 
selective decision-making,
and reproduce experimental results for specific organisms~\cite{Sridhar2021,Moussaid2011,Collignon2016,Gorbonos2024,Ito2024}.

Continuum description has also been used to elucidate the nature of phase transitions
in conventional models of collective motion~\cite{Toner1995,Toner1998,Bertin2009,Bertin2015,Peshkov2012,Patelli2019}.
In particular, the Boltzmann approach, 
which describes 
time evolution of a probability distribution function
by pairwise collision and alignment of agents,
is successfully used to derive hydrodynamic
equations and analyze phase transition~\cite{Bertin2009,Peshkov2012,Bertin2015,Patelli2019}.
However, fundamental aspects of phase transition arising from visual interaction 
is not yet clear.
The difficulty lies in the many-body nature of occlusion, 
where the interaction between two individuals is screened by the other individuals in between.

In this paper, we introduce a Boltzmann approach to 
collective motion induced by non-local visual interaction.
Motivated by the experimental finding~\cite{Jhawar2020}, 
we assume that an agent randomly selects 
a distant neighbor and one at a time,
which has good affinity with the framework 
of the Boltzmann equation. 
We incorporate occlusion by coarse-graining the clouds 
of intervening agents as a density field, which is self-consistently determined by the Boltzmann equation.
This results in an effective pairwise interaction between agents.
In the absence of occlusion, 
the probability to select a neighbor decays with 
a characteristic distance determined 
by the resolution of the eye system. 
We analytically derive the order-disorder transition point, 
and show that it is shifted by the visual screening effect.
Furthermore, we find that the non-locality of interaction
suppresses the discontinuity of the phase transition 
in comparison to local collision models~\cite{Bertin2009,Thuroff2014},
and that the polar order parameter exhibits a mean-field critical behavior
in the weak-advection limit and in a finite system.

\begin{figure}[!b]
\centering
\includegraphics[width=\linewidth]{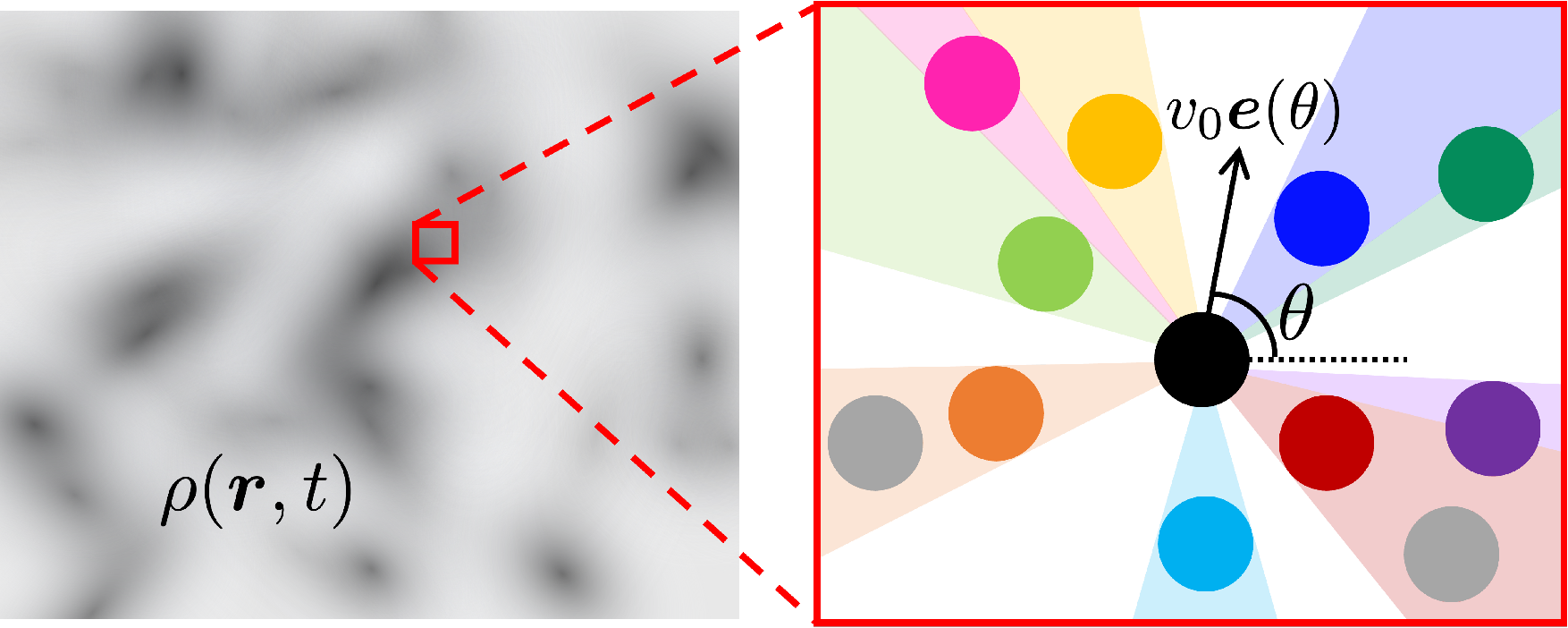}
\caption{
Schematic illustration of the model.
The left panel shows the clouds of agents 
described by the density field $\rho(\bm{r},t)$
in the continuum description,
and the right panel shows the discrete picture.
Each agent has a circular shape and moving with the velocity $v_0\bm{e}(\theta)$.
The center agent (black) perceives the neighbors
and interacts with one of them at a time.
The agents shown in gray are not seen by 
the black agent due to occlusion.
}
\label{schematic}
\end{figure}

%%%%%%%%%%%%%%%%%%%%%%%%%%%%%%%%%%%%%%%%%%%%%%%%%%%%%%%%%%%%
\section{Model}
\subsection{Boltzmann equation}
We consider a two-dimensional model where each agent 
is described by its position $\bm{r}=(x,y)$ and
direction of motion $\bm{e}(\theta)=(\cos\theta,\sin\theta)$.
For simplicity, we assume that 
each agent moves with a constant speed $v_0$ 
and its shape is a circle of diameter $D$
(see Fig.~\ref{schematic}).
We define the probability distribution function 
$f(\bm{r},\theta,t)$ 
and the number density 
of agents
$\rho(\bm{r},t)=\int_{-\pi}^\pi\dd\theta f(\bm{r},\theta,t)$.
The Boltzmann equation describes
time evolution of $f$ via
the self-diffusion term $I_{\mathrm{self}}[f]$ 
and visual interaction term $I_{\mathrm{vis}}[f]$ 
as follows:
\begin{equation}
\label{Boltzmann}
\pdv{f}{t}+v_0\bm{e}(\theta)\cdot\grad f=I_{\mathrm{self}}[f]+I_{\mathrm{vis}}[f],
\end{equation}
\begin{equation}
\label{Iself}
I_{\mathrm{self}}[f]=-sf(\bm{r},\theta,t)+s\int_{-\pi}^\pi\dd\theta'p(\theta-\theta')f(\bm{r},\theta',t),
\end{equation}
\begin{widetext}
\begin{eqnarray}
\label{Ivis}
I_{\mathrm{vis}}[f]&=&-c\int\dd^2\bm{r}'\int_{-\pi}^\pi\dd\theta'G(\bm{r},\bm{r}'-\bm{r},t)\Gamma(\abs{\bm{r}'-\bm{r}},\theta'-\theta)f(\bm{r},\theta,t)f(\bm{r}',\theta',t)\\\nonumber
&&+c\int\dd^2\bm{r}'\int_{-\pi}^\pi\dd\theta_1\int_{-\pi}^\pi\dd\theta_2G(\bm{r},\bm{r}'-\bm{r},t)\Gamma(\abs{\bm{r}'-\bm{r}},\theta_2-\theta_1)f(\bm{r},\theta_1,t)f(\bm{r}',\theta_2,t)\hat{p}(\theta-\vartheta(\theta_1,\theta_2)).
\end{eqnarray}
\end{widetext}

In the self-diffusion integral (Eq.~(\ref{Iself})),
$s$ is the rate of reorientation by the noise, and 
$p(\theta)$ is the probability distribution function of 
angle change:
we use the von Mises distribution $p(\theta)=e^{\kappa\cos\theta}/(2\pi I_0(\kappa))$,
where $I_{n=0,1,\cdots}(\kappa)$ is the modified Bessel function of the first kind
and $\kappa$ is the sharpness parameter~\cite{Collignon2016}.
In Eq.~(\ref{Iself}), 
the first and second term on the right-hand side 
represents the probability of 
transition from $\theta$ to any angle and from $\theta'$ to $\theta$, respectively.

The visual interaction integral (Eq.~(\ref{Ivis}))
describes the pairwise orientational interaction between
the agent at $\bm{r}$ and a neighbor at $\bm{r}'$.
The rate of interaction is 
determined by occlusion by other agents 
and the intrinsic mechanism of visual recognition.
The latter decays with distance 
due to the resolution of eyes~\cite{Pita2015} and
also depends on the relative direction of motion~\cite{Calovi2018}.
Therefore, we introduce the occlusion factor 
$G(\bm{r}, \bm{R}, t)$ 
and the intrinsic factor 
$\Gamma(|\bm{R}|,\psi)$ 
in the interaction kernel, 
where $\bm{R}=\bm{r}'-\bm{r}$ 
and $\psi=\theta'-\theta$ ($\psi=\theta_2-\theta_1$) are the relative position and angle,
respectively.
We will  formulate them in the following paragraphs.
In Eq.~(\ref{Ivis}), $c$ is the reference interaction rate, and
$\hat{p}(\theta)=e^{\hat{\kappa}\cos\theta}/(2\pi I_0(\hat{\kappa}))$ is the probability distribution function of 
angle change for interaction~\cite{SUPprob}.
The first term on the right-hand side
represents the outgoing event in which 
the agent at  ($\bm{r},\theta$) 
interacts with the neighbor at ($\bm{r}',\theta'$)
and is reoriented from $\theta$ to another angle.
The second term represents the incoming event 
where the agent at ($\bm{r},\theta_1$) 
interacts with the neighbor at ($\bm{r}',\theta_2$)
and is reoriented to their average angle
$\theta=\vartheta(\theta_1,\theta_2)=\arg(e^{i\theta_1}+e^{i\theta_2})$~\cite{Bertin2009}.

%%%%%%%%%%%%%%%%%%%%%%%%%%%%%%%%%%%%%%%%%%%%%%%%%%%%%%%%%%
\begin{figure}[!b]
\centering
\includegraphics[width=\linewidth,bb=6 0 587 477]{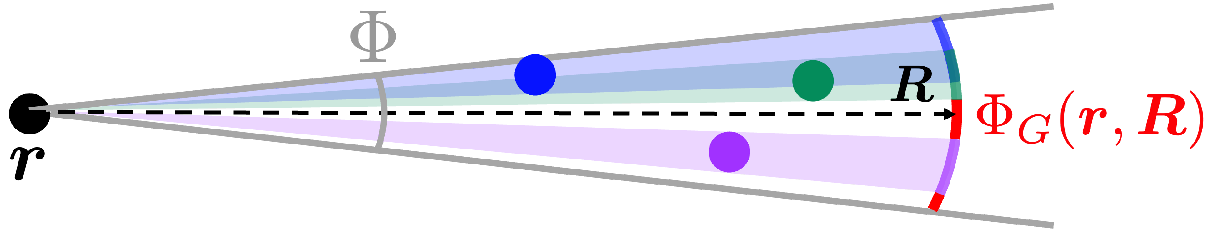}
\caption{
Schematic illustration of $\Phi_G(\bm{r},\bm{R})$.
The agent at the position $\bm{r}$ sees the point 
at the relative position $\bm{R}$ on the line of sight .
The images of the neighbors (the blue, green, and purple agents) in 
the angle $\Phi$ are projected onto the arc, 
in which the non-occupied parts (red) has the total angle $\Phi_G(\bm{r},\bm{R})$.
}
\label{PhiG sch}
\end{figure}

\subsection{occlusion factor}
Now we formulate the occlusion factor 
$G(\bm{r}, \bm{R}, t)$, 
which is a key feature of our model.
The resolution of the eye is limited by the number of ganglion cells in the retina~\cite{Pita2015},
and represented by the azimuthal resolution angle $\Phi$ 
in our two-dimensional model.
Images of the other agents within the angle $\Phi$ centered around the line of sight,
which has the direction $\hat{\bm{R}}=\bm{R}/R$,
are not distinguished from each other. 
We define the angular width $\Phi_G(\bm{r},\bm{R},t)$
of the regions 
that are {\it not} 
occupied by the images of the other agents 
within the distance $R$ and within the angle $\Phi$
(see Fig.~\ref{PhiG sch}).
The occlusion factor $G(\bm{r},\bm{R},t)$ 
is defined as the probability that
a neighbor at $\bm{r}+\bm{R}$ is visible for
the agent at $\bm{r}$.
Therefore, it reads
\begin{equation}
\label{def G}
G(\bm{r},\bm{R},t)=
\frac{\Phi_G(\bm{r},\bm{R},t)}{\Phi}\leq1.
\end{equation}
As the distance is increased
in a given direction $\hat{\bm{R}}$,
more agents come into the line of sight and 
decrease $\Phi_G(\bm{r},\bm{R},t)$.
Let $\sigma(\bm{r}, \bm{R}, t) \dd R$ 
be the fraction of the angular region
occupied by the agents 
in the distance between $R$ and $R+\dd R$
and within the angle $\Phi$.
Given that they are randomly distributed,
the unoccupied angular region $\Phi_G$ is reduced by the same fraction.
Thus we obtain
the differential equation
\begin{equation}
\label{diff occu}
\pdv{R}\Phi_G(\bm{r},\bm{R},t)=
- \sigma(\bm{r}, \bm{R},t)
\Phi_G(\bm{r},\bm{R},t).
\end{equation}
The occupied fraction is related to the local density 
as
\begin{equation}
\label{def sig}
\sigma(\bm{r},\bm{R},t)=
\mathcal{D}(R)\rho(\bm{r}+\bm{R},t),
\end{equation}
where $\mathcal{D}(R)$ is the average arc length occupied by an agent in the visual angular bin $\Phi$. 
This fraction is calculated as the angle occupied by
a single agent $\mathcal{D}/R$ multiplied by
the number of agents in the area $\rho R \Phi\dd R$
and divided by the angular width $\Phi$.
While the angle occupied by a single agent in the field of view of a focal
agent tends to zero in the limit $R \to \infty$, 
the average length of the occupied
visual arc length $\mathcal{D}$ goes to the diameter of the agent $D$. 
At short distances, 
an agent covers the whole angular width $\Phi$ and thus
$\mathcal{D}$ is given by the arc length $R \Phi$.
The formula for intermediate distances is 
derived by straightforward geometric calculation~\cite{SUPocc},
and gives
\begin{align}
\label{D}
\mathcal{D}(R)
=
\left\{ \begin{array}{ll}
R\Phi & [R<R_D], \\
R_D \Phi\left(2-\frac{R_D}{R}\right) & [R> R_D],
\end{array} \right.
\end{align}
where $R_D =D/(2\Phi)$.

The differential equation (\ref{diff occu}) is solved 
under the boundary condition $\Phi_G|_{R\to 0} =\Phi$.
Substituting the solution into Eq.(\ref{def G}), we obtain
\begin{equation}
\label{G}
G(\bm{r},\bm{R},t)=\exp(-\int_0^R\dd R' \mathcal{D}(R')\rho(\bm{r}+\bm{R}',t)),
\end{equation}
The occlusion factor (Eq.~(\ref{G})) gives an
effective pairwise interaction mediated by
the density field $\rho$, which 
is determined by the Boltzmann equation 
in a self-consistent manner.

Note that the coarse-grained description requires
that many neighbors are observed in the resolution angle $\Phi$.
For the mean density $\rho_0$, a neighbor in the typical distance $R \sim1/\sqrt{\rho_0\Phi}$
(note that the number of agents in the distance $R$ and angle $\Phi$ 
is estimated by $n \sim \rho_0 R^2 \Phi/2$)
occupies the angle $\sqrt{\rho_0 \Phi} D$ in the field of view,
which should be much smaller than $\Phi$.
Thus we obtain the condition $\rho_0 \ll \Phi/D^2$ for the density.
Under this condition,  $R \gg R_D$ is satisfied in Eq.~(\ref{D}) for most cases, and 
the occlusion factor $G$ decays with the characteristic 
distance $R_{\rm occ} = 1/(\rho_0 D)\gg R$~\cite{SUPexG}.
Since $\Phi \ll 1$, the above condition also ensures that
the area fraction $A=\rho_0\pi\leri{D/2}^2$ is much smaller than unity,
which is required for the excluded-volume interaction to be negligible.

%%%%%%%%%%%%%%%%%%%%%%%%%%%%%%%%%%%%%%%%%%%%%%%%%%%%%%%%%%%
\subsection{intrinsic factor}
The intrinsic factor $\Gamma(R,\psi)$ depends
on the relative distance $R$ and the relative angle $\psi$.
We assume that it can be factorized as 
$\Gamma(R,\psi)=B(R)K(\psi)$.
The factor $B(R)$ represents the possibility of visual recognition of the direction of motion of a neighbor.
The probability of visual recognition
decreases as the relative distance increases.
We then define
\begin{equation}
B(R)=\exp(-\frac{R^2}{2R_0^2}),
\label{GammaB}
\end{equation}
and
$R_0$ is the characteristic distance of the visual recognition.
In fact, the magnitude of orientational interaction of some species of fish
is found to obey the Gaussian function~\cite{Calovi2018}.

The factor $K(\psi)$ represents the probability of reaction to a neighbor at the relative angle $\psi$.
The detailed study for fish~\cite{Calovi2018} shows
\begin{equation}
K(\psi)=\frac{\abs{\sin\psi}\{1+a\cos(2\psi)\}}{\sqrt{(2-2a+a^2)/4}},
\end{equation}
where 
the essential term is $\abs{\sin\psi}$ and the $\cos(2\psi)$ term
is introduced for fitting with the experimental data~\cite{Calovi2018}.
The normalization $\frac{1}{2\pi}\int_{-\pi}^\pi\dd\psi K^2(\psi)=1$ is assumed.
In this paper, we use
\begin{equation}
K(\psi)=\sqrt{2}\abs{\sin\psi}
\label{GammaK}
\end{equation}
by setting $a=0$ for simplicity
and in order to make our model more applicable to various organisms~\cite{SUPintrinsic}.

%%%%%%%%%%%%%%%%%%%%%%%%%%%%%%%%%%%%%%%%%%%%%%%%%%%%%%%%%%%%
\section{Results}
\subsection{linear stability}
As we increase the interaction rate $c$,
the uniform disordered state $f=\overline{f_0}$
(const.) becomes unstable.
In order to obtain the transition point $c=c_{\mathrm{tr}}$,
we performed a linear stability analysis of the Boltzmann equation (see Appendix 5).
We add the perturbation $\delta f(\bm{r},\theta,t)$ to $\rho_0$,
and define its Fourier component $\delta f_{k}(\bm{q})$ and 
the complex damping rate $\Lambda_k(\bm{q})$ via:
\begin{equation}
\label{}
\delta f(\bm{r},\theta,t)=\int\dd^2\bm{q}\sum_{k} 
\delta f_k(\bm{q})e^{i(\bm{q}\cdot\bm{r}+k\theta)}e^{-\Lambda_k(\bm{q})t},
\end{equation}
where $k=0, \pm 1, \pm 2, \cdots$ is the index of the Fourier modes in the angular domain.
From the stability analysis,
we find that only the damping rate of the mode $k=\pm 1$ can become negative.
Their damping rate is given by
\begin{eqnarray}
\label{ReLamb1}
\Re\Lambda_{\pm1}(\bm{q})&=&s\leri{1-2\pi p_1}\\\nonumber
&&-4\sqrt{2}c\mathcal{I}\overline{f_0}\left\{\frac{2}{3}(2\pi\hat{p}_1)\leri{1+\hat{\mathcal{I}}(\bm{q})}-1\right\},
\end{eqnarray}
where $p_1$ and $\hat{p}_1$ are the Fourier components of $p(\theta)$ and $\hat{p}(\theta)$ for $k=1$,  respectively, and
\begin{eqnarray}
\label{}
\mathcal{I}\overline{f_0}&=&\frac{1-\exp(-\frac{\rho_0R_D^2}{2}\leri{\Phi+\frac{1}{\rho_0R_0^2}})}{\Phi+\frac{1}{\rho_0R_0^2}}\\\nonumber
&&+\rho_0R_D^2\int_1^\infty\dd \xi\xi e^{-\frac{R_D^2}{2R_0^2}\xi^2}\leri{e^{\frac{3}{2}}\xi e^{-2\xi}}^{\rho_0\Phi R_D^2}.
\end{eqnarray}
is a function of the dimensionless quantities $\rho_0R_D^2$, $\rho_0R_0^2$, and $\Phi$.
In Eq.~(\ref{ReLamb1}),
the wavenumber-dependent function $\hat{\mathcal{I}}(\bm{q})$ 
satisfies $-1<\hat{\mathcal{I}}(\bm{q})\leq1$, and
has the single maximum $\hat{\mathcal{I}}(\bm{0})=1$.
Thus the transition point is determined by $\Re\Lambda_1(\bm{0})=0$.
In the no-occlusion case obtained in the point particle limit $D=0$, 
we get $\hat{\mathcal{I}}(\bm{q})=e^{-q^2R_0^2/2}$.
In the linear stability analysis, 
we neglected the advection term in Eq.(\ref{Boltzmann}),
which does not affect the transition point~\cite{Bertin2009,Patelli2019}
(see also Appendix 5).

\begin{figure}[!t]
\centering
\includegraphics[width=\linewidth]{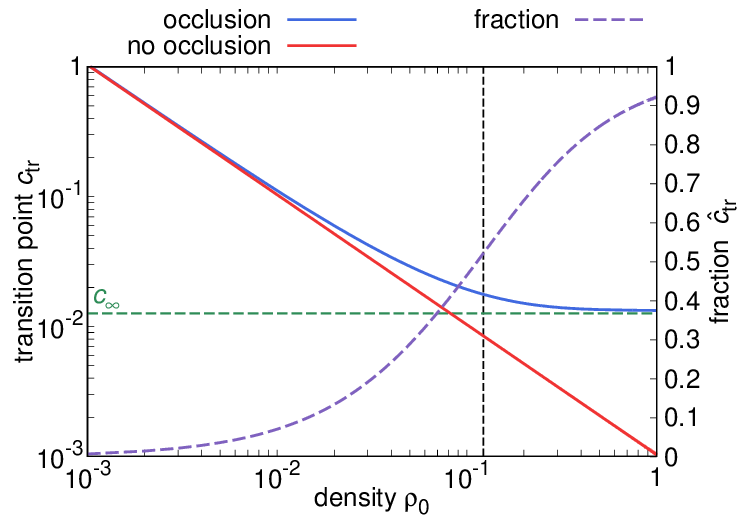}
\caption{
The transition point $c_{\mathrm{tr}}$ as a function of the density $\rho_0$.
We set the biologically reasonable parameter values
$s=2,~\kappa=6,~\hat{\kappa}=20,~\Phi=7^\circ,~R_0=10$~\cite{Pita2015,Calovi2018,Collignon2016,Bainbridge1958,Akanyeti2017}
which are rescaled by the diameter $D=1$ body length and 1 sec~\cite{SUPparameter}.
The blue solid line shows $c_{\mathrm{tr}}$ (Eq.~(\ref{ctr})),
and reaches the horizontal green dashed line $c_\infty$ in high density limit.
Note that the theory is valid only for $\rho_0 \ll \Phi/D^2$, and 
the black dashed line marks $\rho_0 = \Phi/D^2\simeq0.12$.
The red solid line corresponds to the no occlusion case $c_{\mathrm{tr}}|_{D=0}$.
The purple dashed line is the fraction $\hat{c}_{\mathrm{tr}}$ to the right vertical axis.
}
\label{rho ctr}
\end{figure}

%%%%%%%%%%%%%%%%%%%%%%%%%%%%%%%%%%%%%%%%%%%%%%%%%%%%%%%%%%%%
\subsection{transition point}
We then obtain the transition point 
\begin{equation}
\label{ctr}
c_{\mathrm{tr}}=\frac{s}{4\sqrt{2}\mathcal{I}\overline{f_0}}\frac{1-2\pi p_1}
{\frac{4}{3}(2\pi\hat{p}_1)-1}:=\frac{c_0}{\mathcal{I}\overline{f_0}},
\end{equation}
where $c_0$ increases when $s$ increases, or when $\kappa$ or $\hat{\kappa}$ decreases,
which means that the transition is hindered by the orientational noise.
Note that $2\pi\hat{p}_1$ must be larger than $3/4$ such that $c_0>0$.
As shown in Fig.~\ref{rho ctr},
$c_{\mathrm{tr}}$ decreases as $\rho_0$ increases,
which meets the expectation that a higher density of agents 
induces stronger alignment.
It is notable that $c_{\mathrm{tr}}$ is non-zero in the limit $R_0 \to \infty$, 
where the interaction decays only due to occlusion.
In contrast, for $D=0$,
we obtain $c_{\mathrm{tr}}|_{D=0}=c_0/(\rho_0R_0^2)$, 
which vanishes
in the limit $R_0\to \infty$.

In Fig.~\ref{rho ctr}, we also show
the fraction $\hat{c}_{\mathrm{tr}}=(c_{\mathrm{tr}}-c_{\mathrm{tr}}|_{D=0})/c_{\mathrm{tr}}$.
As a function of the density,  it shows a sizable increase before reaching 
the limit of validity of the model $\rho_0 = \Phi/D^2$,
and then converges to unity as $\rho_0\to\infty$.

%%%%%%%%%%%%%%%%%%%%%%%%%%%%%%%%%%%%%%%%%%%%%%%%%%%%%%%%%%%%
\section{Discussions}

We summarize the new aspects of our analytical treatment in comparison with previous Boltzmann-type models of active matter as follows:
(i)
We have modeled the visual interaction by the Boltzmann approach for the first time. 
(ii)
The screening effect (occlusion) is represented by a density field,
which is self-consistently determined from the distribution function.
This has enabled the effective description of many-body interaction 
in the Boltzmann approach. 
Except for occlusion, the interaction is assumed to be pairwise, which
is justified by the experimental observation in Ref.~\cite{Jhawar2020}.
We incorporated the non-local visual interaction 
to the Boltzmann equation as the effective pairwise interaction.
Our model shows that the visual screening effect 
raises the transition threshold for the interaction rate $c$,
which becomes non-zero even in the limits of high density 
or infinite range of the intrinsic interaction.
This result is in stark contrast to those of the local-interaction
models that assume point-like particles~\cite{Bertin2009,Bertin2015}.
In those models, the transition threshold was determined by 
hydrodynamic equations that are obtained by truncating 
the hierarchical equations for the angular Fourier modes 
$f_k(\bm{r},t)= (2\pi)^{-1}\int_{-\pi}^\pi f(\bm{r},\theta,t) e^{-ik\theta} d\theta$.
Our results reproduce those of the previous models of polar particles with ferromagnetic 
interactions by taking the local limit
$R_0 \to 0$ with $cR_0^2$ kept constant.
In this limit, the interaction kernel $I_{\mathrm{vis}}[f]$
converges to the delta function $\delta(\bm{r}'-\bm{r})$~\cite{SUPnonlocal},
and the linear growth rate $-\Re\Lambda_1(\bm{0})$ becomes 
identical to $\mu$ of Eq.~(31) in Ref.~\cite{Bertin2009}.

The damping rate $\Re \Lambda_1$ also contains the information 
of the kind of the disorder-order phase transition.
In previous local interaction models,
instability of the modes with $\bm{q}\neq \bm{0}$ induce
phase coexistence in the form of traveling bands,
which is the origin of the discontinuous transition~\cite{Bertin2009}.
Therefore, the transition is continuous 
if the $\bm{q}\neq \bm{0}$ modes are linearly stable
at the transition point.
This holds if the damping rate has no density-dependence, 
which is the case for metric-free interactions~\cite{Peshkov2012}.
On the other hand, the  transition becomes discontinuous
if the damping rate depends on the density~\cite{Peshkov2012}.
In our model, we have $\pdv*{\Lambda_1}{\rho_0}\neq0$
and thus our system shows a discontinuous transition. 
The non-locality of the interaction does not affect the kind of the phase transition.
This is reasonable if we consider that
the Vicsek model~\cite{Vicsek1995} has a finite interaction range
and shows a discontinuous transition~\cite{chate2008collective}.
Active nematics with non-local repulsion also shows
a discontinuous order-disorder transition~\cite{Patelli2019}.
It is also worth mentioning that
the visual interaction with occlusion in high density regions
resembles metric-free interactions with the nearest topological neighbors,
which may render the transition density independent. 
However, recent works~\cite{Martin2021,Rahmani2021} found emergence 
of traveling bands even for purely topological interactions, and proposed 
different mechanisms that induce effective density-order couplings.
It suggests the possibility that 
the order-disorder transition in aligning active matter systems
may be always discontinuous due to implicit density dependence.

Let us now consider the model's behavior above the transition point.
The advection term in the Boltzmann equation (\ref{Boltzmann})
introduces couplings between $f_{k}$ and $f_{k\pm 1}$ in the mode expansion,
which introduces an infinite-dimensional matrix in the linear stability analysis. 
However, in our model, only the mode with $k=\pm 1$ is linearly unstable
at the transition point in the non-advective limit ($v_0=0$).
It means that there is a finite window of $c$ 
above the transition point where the damping rate for $f_{k\neq \pm1}$ 
remains positive. In this window, contributions 
to $f_{|k|> 1}$ 
%($f_{k< -1}$)
by the advective couplings are 
$\mathcal{O}\leri{v_0^{|k|-1} f_{\pm 1}}$, 
%${\cal O}(v_0^{-k+1} f_{-1})$) (resp.),
which we can neglect compared to non-advective terms for sufficiently small values of $v_0$.
Thus we can use the damping rate $\Lambda_k(\bm{q})$ obtained 
in the non-advective limit ($v_0=0$).

In our model, the wavenumber-dependence of the damping rates comes
through $\hat{\mathcal{I}}(\bm{q})$ in Eq.~(\ref{ReLamb1}).
Since $\hat{\mathcal{I}}(\bm{q})$ is a decreasing function of $q$ and $R_0$,
only long-wavelength modes are destabilized in the vicinity of the transition point,
and the unstable range of $q$ gets narrower 
for a larger interaction range.
In particular, when we consider a finite two-dimensional system
with the periodic boundary condition, 
the wavenumber is discretized and 
there is a finite window of $c$ above $c_{\rm tr}$
where only the $\bm{q}=\bm{0}$ mode becomes unstable.
In this case, the transition becomes continuous and we can calculate
the polar order parameter defined by
$P= \left|
\int_{-\pi}^\pi\dd\theta\bm{e}(\theta)\langle f(\bm{r},\theta,t)\rangle \right|
/\rho_0 = |\langle f_1 \rangle |/f_0$,
where the brackets mean spatiotemporal and ensemble averages.
By truncating the hierarchical equations for $f_k$,
we obtain the scaling law 
$P=P_0(c_{\mathrm{tr}})(c-c_\mathrm{tr})^{1/2}$
~\cite{SUPscaling}.
As shown in Fig.~S3,
the prefactor $P_0(c_{\mathrm{tr}})$ is larger for smaller $c_{\mathrm{tr}}$,
which corresponds to large $\rho_0$ and/or large $\hat{\kappa}$.

Finally,
we compare our result with the behavior of real organisms,
using the examples of fish in a shallow tank showing two-dimensional collective motion.
A phase transition is observed for tilapia (\textit{O. niloticus} L.)
when the number density increases above a threshold~\cite{Becco2006}, 
which is in the range $\rho_0 \in [0.01, 0.1]$ in our model.
(In Ref.~[39], the number density at the transition point is 472 fishes/m$^2$ 
and the body length of specimens is 11-15 mm.
Using the body length as the unit of length, 
we obtain the dimensionless density $\rho_0$=0.057-0.106 fishes/BL$^2$.
Considering the slenderness of the body of real fish, 
which reduces occupancy in the visual field compared to the circular fish in our model,
we estimate the effective density to be in the range $\rho_{0}\in[0.01,0.1]$.)
For $\rho_0=0.01$, the fraction $\hat{c}_{\mathrm{tr}}$ is about 0.1,
and thus occlusion has small but non-negligible effect. 
However, the experimental density is closer to the threshold $\rho_0=\Phi/D^2
\simeq 0.12$,
and our results are not directly applicable to the experiment.
For near and above the threshold density where a few agents occupy the visual field,
we can consider a hybrid model:
we treat the agents as discrete self-propelled particles in the near field,
and use continuum description with the probability distribution function 
at the far field.
This will reduce the computational cost for treating many fish in the far field,
and to be addressed in future work.
A quantitative study of occlusion in different species of fish (or robot fish) 
will be necessary to verify the results of our model.  
On the other hand, as the number density increases,
golden shiner (\textit{N. crysoleucas}) shows rotating state~\cite{Tunstrom2013},
and cichlid (\textit{E. suratensis}) exhibits many oriented sub-clusters
which is not aligned as a whole cluster~\cite{Jhawar2020}.
There are many aspects of interactions 
that are considered in previous agent-based models 
but not in our model.
The emergence of a rotating state is facilitated
by introducing the dead angle~\cite{Costanzo2018}
and the wall~\cite{Gautrais2009}, and
considering three-dimensional motion~\cite{Ito2022}.
The polar order is also disturbed by
anisotropic attraction and repulsion~\cite{Katz2011},
the change of speed by interaction~\cite{Bastien2020,Ito2024},
and non-circular shapes of an agent~\cite{Pearce2014}.
The effects of these interesting extensions
by modifying the spatial integral kernel in the Boltzmann equation
are the subject of future work.

%%%%%%%%%%%%%%%%%%%%%%%%%%%%%%%%%%%%%%%%%%%%%%%%%%%%%%%%%%%
\begin{acknowledgments}
We acknowledge financial support by JSPS KAKENHI Grant No. 23KJ0171 to S.I.
and support by a research environment of Tohoku University,
Division for Interdisciplinary Advanced Research and Education to S.I.
\end{acknowledgments}
%%%%%%%%%%%%%%%%%%%%%%%%%%%%%%%%%%%%%%%%%%%%%%%%%%%%%%%%%%%
\begin{widetext}
\appendix*
\section{Stability analysis of the Boltzmann equation}
\subsection{Fourier components of the distribution function}
Here, we perform the linear stability analysis for Eq.~(\ref{Boltzmann}).
The aim is the derivation of the transition point.

We consider the perturbation from the uniform distribution $\overline{f_0}$:
\begin{equation}
\label{}
f(\bm{r},\theta,t)=\overline{f_0}+\delta f(\bm{r},\theta,t).
\end{equation}
The density is
\begin{equation}
\label{}
\rho(\bm{r},t)=2\pi \overline{f_0}+\int_{-\pi}^\pi\dd\theta\delta f(\bm{r},\theta,t)=\rho_0+\delta\rho(\bm{r},t),
\end{equation}
and the occlusion factor reads
\begin{equation}
\label{}
G(\bm{r},\bm{R},t)\simeq\exp(-\rho_0\int_0^R\dd R'\mathcal{D}(R'))\times\leri{1-\int_0^R\dd R'\mathcal{D}(R')\delta\rho(\bm{r}+\bm{R}',t)}:=G_0(R)\{1-\delta G(\bm{r},\bm{R},t)\}.
\end{equation}

The Boltzmann equation (Eq.~(\ref{Boltzmann})) becomes 
\begin{eqnarray}
\label{Boltzmann df}
\pdv{\delta f(\bm{r},\theta,t)}{t}&=&-v_0\bm{e}(\theta)\cdot\grad\delta f(\bm{r},\theta,t)\\\nonumber
&&-s\delta f(\bm{r},\theta,t)+s\int_{-\pi}^\pi\dd\theta'p(\theta-\theta')\delta f(\bm{r},\theta',t)\\\nonumber
&&-c\mathcal{I}\overline{f_0}\int_{-\pi}^\pi\dd\theta'K(\theta'-\theta)\\\nonumber
&&~~~~~\times\left\{\overline{f_0}\leri{1-\frac{\delta\mathcal{I}(\bm{r},t)}{\mathcal{I}}}+\delta f(\bm{r},\theta,t)+\frac{1}{\mathcal{I}}\int\dd^2\bm{r}'G_0(\abs{\bm{r}'-\bm{r}})B(\abs{\bm{r}'-\bm{r}})\delta f(\bm{r}',\theta',t)\right\}\\\nonumber
&&+c\mathcal{I}\overline{f_0}\int_{-\pi}^\pi\dd\theta_1\int_{-\pi}^\pi\dd\theta_2K(\theta_2-\theta_1)\hat{p}(\theta-\vartheta(\theta_1,\theta_2))\\\nonumber
&&~~~~~\times\left\{\overline{f_0}\leri{1-\frac{\delta\mathcal{I}(\bm{r},t)}{\mathcal{I}}}+\delta f(\bm{r},\theta_2,t)+\frac{1}{\mathcal{I}}\int\dd^2\bm{r}'G_0(\abs{\bm{r}'-\bm{r}})B(\abs{\bm{r}'-\bm{r}})\delta f(\bm{r}',\theta_2,t)\right\}.
\end{eqnarray}
where we use $K(-\psi)=K(\psi)$ and $\vartheta(\theta_1,\theta_2)=\vartheta(\theta_2,\theta_1)$ and define
\begin{equation}
\label{}
\mathcal{I}=\int\dd^2\bm{R}G_0(R)B(R),~\delta\mathcal{I}(\bm{r},t)=\int\dd^2\bm{R}G_0(R)B(R)\delta G(\bm{r},\bm{R},t).
\end{equation}

We define the Fourier transform for a function $F(\bm{r},\theta)$:
\begin{equation}
\label{}
F(\bm{r},\theta)=\int\dd^2\bm{q}\sum_kF_k(\bm{q})e^{i(\bm{q}\cdot\bm{r}+k\theta)},~F_k(\bm{q})=\frac{1}{(2\pi)^3}\int\dd^2\bm{r}\int_{-\pi}^\pi\dd\theta F(\bm{r},\theta)e^{-i(\bm{q}\cdot\bm{r}+k\theta)},
\end{equation}
where $k=0,\pm1,\pm2,\cdots$ is the discrete wavenumber of the angle Fourier transform,
and $\bm{q}$ is the wavenumber vector of the spatial Fourier transform.
We introduce the damping rate $\Lambda_k(\bm{q})$ for the Fourier component via
\begin{equation}
\label{}
\delta f(\bm{r},\theta,t)=\int\dd^2\bm{q}\sum_k\delta f_k(\bm{q})e^{i(\bm{q}\cdot\bm{r}+k\theta)}e^{-\Lambda_k(\bm{q})t}:=\int\dd^2\bm{q}\sum_k\delta f_k(\bm{q},t)e^{i(\bm{q}\cdot\bm{r}+k\theta)}.
\end{equation}
(The index of zero represents $k=0$, except for $\overline{f_0}$, $\rho_0$, and the parameters).

%%%%%%%%%%%%%%%%%%%%%%%%%%%%%%%%%%%%%%%%%%%%%%%%%%%%%%%%%%%%%%
\subsection{Terms not including $\delta f$}
We confirm that the terms which do not include explicitly $\delta f$ are canceled out.
In other wards,
\begin{equation}
\label{Ndf 1}
-c\mathcal{I}\overline{f_0}^2\leri{1-\frac{\delta\mathcal{I}(\bm{r},t)}{\mathcal{I}}}\int_{-\pi}^\pi\dd\theta'K(\theta'-\theta)=-2\pi c\mathcal{I}\overline{f_0}^2\leri{1-\frac{\delta\mathcal{I}(\bm{r},t)}{\mathcal{I}}}K_0
\end{equation}
and
\begin{equation}
\label{Ndf 2}
c\mathcal{I}\overline{f_0}^2\leri{1-\frac{\delta\mathcal{I}(\bm{r},t)}{\mathcal{I}}}\int_{-\pi}^\pi\dd\theta_1\int_{-\pi}^\pi\dd\theta_2K(\theta_2-\theta_1)\hat{p}(\theta-\vartheta(\theta_1,\theta_2))
\end{equation}
are canceled out as follows.

The angle $\vartheta(\theta_1,\theta_2)$ is $\arg(e^{i\theta_1}+e^{i\theta_2})$ in the Boltzmann equation,
but we must consider the absolute value of $e^{i\theta_1}+e^{i\theta_2}$
for accurate treatment of $\arg(\circ)$ in the Fourier transform for the calculation of
$\int_{-\pi}^\pi\dd\theta_1\int_{-\pi}^\pi\dd\theta_2K(\theta_2-\theta_1)\hat{p}(\theta-\vartheta(\theta_1,\theta_2))$.
We therefore redefine 
\begin{equation}
\label{}
\vartheta(\theta_1,\theta_2)=\arg\leri{\frac{e^{i\theta_1}+e^{i\theta_2}}{\abs{2\cos(\frac{\theta_2-\theta_1}{2})}}},
\end{equation}
and the integral is
\begin{eqnarray}
\label{}
&&\int_{-\pi}^\pi\dd\theta_1\int_{-\pi}^\pi\dd\theta_2K(\theta_2-\theta_1)\hat{p}(\theta-\vartheta(\theta_1,\theta_2))\\\nonumber
&=&\int_{-\pi}^\pi\dd\theta_1\int_{-\pi}^\pi\dd\theta_2\sum_kK_ke^{ik(\theta_2-\theta_1)}\sum_{k'}\hat{p}_{k'}e^{ik'(\theta-\vartheta(\theta_1,\theta_2))}\\\nonumber
&=&\int_{-\pi}^\pi\dd\theta_1\int_{-\pi}^\pi\dd\theta_2\sum_{k}K_ke^{ik(\theta_2-\theta_1)}\sum_{k'}\hat{p}_{k'}e^{ik'\theta}\leri{\frac{e^{i\theta_1}+e^{i\theta_2}}{\abs{2\cos(\frac{\theta_2-\theta_1}{2})}}}^{-k'}\\\nonumber
&=&\int_{-\pi}^\pi\dd\theta_1\int_{-\pi}^\pi\dd\theta_2\sum_{k}K_ke^{ik(\theta_2-\theta_1)}\sum_{k'}\hat{p}_{k'}e^{ik'\theta}\left\{\sgn{\cos(\frac{\theta_2-\theta_1}{2})}\right\}^{k'}e^{-i\frac{k'}{2}(\theta_1+\theta_2)}.
\end{eqnarray}
We convert the variables from $(\theta_1,\theta_2)$ to
$\leri{\Theta=\frac{\theta_1+\theta_2}{2},~\Psi=\frac{\theta_2-\theta_1}{2}}$
and the Jacobian is
$\dd\theta_1\dd\theta_2=2\dd\Theta\dd\Psi$.
Then the above becomes
\begin{eqnarray}
\label{Int29}
&&2\sum_{k}\sum_{k'}K_k\hat{p}_{k'}e^{ik'\theta}\int_{-\pi}^\pi\dd\Psi e^{2ik\Psi}\left\{\sgn{\cos\Psi}\right\}^{k'}\int_{-\pi+\abs{\Psi}}^{\pi-\abs{\Psi}}\dd\Theta e^{-ik'\Theta}\\\nonumber
&=&4\sum_{k}K_k\hat{p}_{0}\int_{-\pi}^\pi\dd\Psi e^{2ik\Psi}(\pi-\abs{\Psi})-4\sum_{k}\sum_{k'\neq0}K_k\hat{p}_{k'}e^{ik'\theta}\int_{-\pi}^\pi\dd\Psi e^{2ik\Psi}\left\{-\sgn{\cos\Psi}\right\}^{k'}\frac{\sin(k'\abs{\Psi})}{k'}.
\end{eqnarray}
In the case of $k\neq0$, the first integral on the right hand side of Eq.~(\ref{Int29}) is zero.
Therefore, the first integral becomes
\begin{equation}
\label{}
4\sum_{k}K_k\hat{p}_{0}\int_{-\pi}^\pi\dd\Psi e^{2ik\Psi}(\pi-\abs{\Psi})=4K_0\hat{p}_{0}\int_{-\pi}^\pi\dd\Psi(\pi-\abs{\Psi})=4\pi^2K_0\hat{p}_{0}=2\pi K_0,
\end{equation}
where $2\pi\hat{p}_{0}=\int_{-\pi}^\pi\dd\theta\hat{p}(\theta)=1$ because $\hat{p}(\theta)$ is the probability distribution. 
On the other hand, the second integral is
\begin{eqnarray}
\label{zero sin}
&&-4\sum_{k}\sum_{k'\neq0}K_k\hat{p}_{k'}e^{ik'\theta}\int_{-\pi}^\pi\dd\Psi e^{2ik\Psi}\left\{-\sgn{\cos\Psi}\right\}^{k'}\frac{\sin(k'\abs{\Psi})}{k'}\\\nonumber
&=&-8\sum_{k}\sum_{k'\neq0}K_k\hat{p}_{k'}\frac{e^{ik'\theta}}{k'}\int_0^\pi\dd\Psi\cos(2k\Psi)\left\{-\sgn{\cos\Psi}\right\}^{k'}\sin(k'\Psi)\\\nonumber
&=&-8\sum_{k}\sum_{k'\neq0}K_k\hat{p}_{k'}\frac{e^{ik'\theta}}{k'}\int_0^\pi\dd\Psi\left\{-\sgn{\cos\Psi}\right\}^{k'}\frac{\sin((k'+2k)\Psi)+\sin((k'-2k)\Psi)}{2}\\\nonumber
&=&-4\sum_{k}\sum_{k'\neq0,\pm2k}K_k\hat{p}_{k'}\frac{e^{ik'\theta}}{k'}\int_0^\pi\dd\Psi\left\{-\sgn{\cos\Psi}\right\}^{k'}\{\sin((k'+2k)\Psi)+\sin((k'-2k)\Psi)\}\\\nonumber
&=&-4\sum_{k}\sum_{k'\neq0,\pm2k}K_k\hat{p}_{k'}\frac{e^{ik'\theta}}{k'}((-1)^{k'}-1)\leri{\frac{\cos(\frac{\pi}{2}(k'+2k))}{k'+2k}+\frac{\cos(\frac{\pi}{2}(k'-2k))}{k'-2k}}\\\nonumber
&=&0,
\end{eqnarray}
because $\cos(\frac{\pi}{2}(k'\pm2k))=0$ when $(-1)^{k'}-1\neq0$ ($k'$ is odd).

Thus, Eq.~(\ref{Ndf 2}) reads
\begin{equation}
\label{}
c\mathcal{I}\overline{f_0}^2\leri{1-\frac{\delta\mathcal{I}(\bm{r},t)}{\mathcal{I}}}\int_{-\pi}^\pi\dd\theta_1\int_{-\pi}^\pi\dd\theta_2K(\theta_2-\theta_1)\hat{p}(\theta-\vartheta(\theta_1,\theta_2))=2\pi c\mathcal{I}\overline{f_0}^2\leri{1-\frac{\delta\mathcal{I}(\bm{r},t)}{\mathcal{I}}}K_0,
\end{equation}
which cancels out with Eq.~(\ref{Ndf 1}).
Therefore, the Boltzmann equation (Eq.~(\ref{Boltzmann df})) becomes
\begin{eqnarray}
\label{Boltzeq2}
\pdv{\delta f(\bm{r},\theta,t)}{t}&=&-v_0\bm{e}(\theta)\cdot\grad\delta f(\bm{r},\theta,t)\\\nonumber
&&-s\delta f(\bm{r},\theta,t)+s\int_{-\pi}^\pi\dd\theta'p(\theta-\theta')\delta f(\bm{r},\theta',t)\\\nonumber
&&-c\mathcal{I}\overline{f_0}\int_{-\pi}^\pi\dd\theta'K(\theta'-\theta)\\\nonumber
&&~~~~~\times\left\{\delta f(\bm{r},\theta,t)+\frac{1}{\mathcal{I}}\int\dd^2\bm{r}'G_0(\abs{\bm{r}'-\bm{r}})B(\abs{\bm{r}'-\bm{r}})\delta f(\bm{r}',\theta',t)\right\}\\\nonumber
&&+c\mathcal{I}\overline{f_0}\int_{-\pi}^\pi\dd\theta_1\int_{-\pi}^\pi\dd\theta_2K(\theta_2-\theta_1)\hat{p}(\theta-\vartheta(\theta_1,\theta_2))\\\nonumber
&&~~~~~\times\left\{\delta f(\bm{r},\theta_2,t)+\frac{1}{\mathcal{I}}\int\dd^2\bm{r}'G_0(\abs{\bm{r}'-\bm{r}})B(\abs{\bm{r}'-\bm{r}})\delta f(\bm{r}',\theta_2,t)\right\}.
\end{eqnarray}

%%%%%%%%%%%%%%%%%%%%%%%%%%%%%%%%%%%%%%%%%%%%%%%%%%%%%%%%%%%%%%
\subsection{Spatial integrals}

Here we calculate the integral $\mathcal{I}\overline{f_0}$.
%Here we perform the spatial integral up to the infinite distance.
\begin{equation}
\label{}
\mathcal{I}\overline{f_0}=\overline{f_0}\int\dd^2\bm{R}B(R)G_0(R)=\rho_0\int_0^\infty\dd RRe^{-\frac{R^2}{2R_0^2}}\exp(-\rho_0\int_0^R\dd R'\mathcal{D}(R')).
\end{equation}
By using Eq.~(\ref{D}), we get the integral in the exponential function:
\begin{align}
\label{}
\int_0^R\dd R'\mathcal{D}(R')= \left\{ \begin{array}{ll}
\frac{R^2}{2}\Phi & \left[R<R_D\right], \\
&\\
R_D^2\Phi\leri{-\frac{3}{2}+2\frac{R}{R_D}+\ln(\frac{R_D}{R})} & \left[R>R_D\right],\
\end{array} \right.
\end{align}
and then
\begin{equation}
\label{If0 D0}
\mathcal{I}\overline{f_0}=\rho_0\int_0^{R_D}\dd RRe^{-\rho_0\leri{\Phi+\frac{1}{\rho_0R_0^2}}\frac{R^2}{2}}+\rho_0\int_{R_D}^\infty\dd RRe^{-\frac{R^2}{2R_0^2}}e^{-\rho_0R_D^2\Phi\left\{-\frac{3}{2}+2\frac{R}{R_D}+\ln(\frac{R_D}{R})\right\}}.
\end{equation}
The first integral is calculated as
\begin{equation}
\label{}
\rho_0\int_0^{R_D}\dd RRe^{-\rho_0\leri{\Phi+\frac{1}{\rho_0R_0^2}}\frac{R^2}{2}}=\frac{1-\exp(-\frac{\rho_0R_D^2}{2}\leri{\Phi+\frac{1}{\rho_0R_0^2}})}{\Phi+\frac{1}{\rho_0R_0^2}}.
\end{equation}
This is zero for $\rho_0=0$, and approaches asymptotically $\frac{1}{\Phi}\leri{1-\frac{1}{\rho_0\Phi R_0^2}}$ for $\rho_0\rightarrow\infty$.
On the other hand, the second integral is
\begin{equation}
\label{}
\rho_0\int_{R_D}^\infty\dd RRe^{-\frac{R^2}{2R_0^2}}e^{-\rho_0\Phi R_D^2\left\{-\frac{3}{2}+2\frac{R}{R_D}+\ln(\frac{R_D}{R})\right\}}=\rho_0R_D^2\int_1^\infty\dd \xi\xi e^{-\frac{R_D^2}{2R_0^2}\xi^2}\leri{e^{\frac{3}{2}}\xi e^{-2\xi}}^{\rho_0\Phi R_D^2},~\xi=\frac{R}{R_D}.
\end{equation}
It is difficult to perform this integral analytically, but we can estimate its value.
$e^{\frac{3}{2}}\xi e^{-2\xi}$ takes $e^{-\frac{1}{2}}$ as the maximum value at $\xi=1$ in the interval of integral,
and then 
\begin{equation}
\label{}
\rho_0R_D^2\int_1^\infty\dd \xi\xi e^{-\frac{R_D^2}{2R_0^2}\xi^2}\leri{e^{\frac{3}{2}}\xi e^{-2\xi}}^{\rho_0\Phi R_D^2}<\rho_0R_D^2e^{-\frac{1}{2}\rho_0\Phi R_D^2}\int_1^\infty\dd \xi\xi e^{-\frac{R_D^2}{2R_0^2}\xi^2}=\rho_0R_0^2e^{-\frac{1}{2}\rho_0\Phi R_D^2}e^{-\frac{R_D^2}{2R_0^2}}.
\end{equation}
This is zero for $\rho_0=0$, and converges to zero exponentially at $\rho_0\rightarrow\infty$.
Therefore, $\mathcal{I}\overline{f_0}$ takes a finite value.
In the point particle limit $R_D=0$ (in other words, the non occlusion case $G=1$),
we can calculate $\mathcal{I}$ as the accurate form 
\begin{equation}
\label{point If}
\mathcal{I}\overline{f_0}=\rho_0\int_0^\infty\dd RRe^{-\frac{R^2}{2R_0^2}}=\rho_0R_0^2.
\end{equation}

Next, we calculate the integral
\begin{equation}
\label{}
\frac{1}{\mathcal{I}}\int\dd^2\bm{r}'G_0(\abs{\bm{r}'-\bm{r}})B(\abs{\bm{r}'-\bm{r}})e^{i\bm{q}\cdot\bm{r}'}=\frac{e^{i\bm{q}\cdot\bm{r}}}{\mathcal{I}}\int\dd^2\bm{R}G_0(R)B(R)e^{i\bm{q}\cdot\bm{R}},
\end{equation}
which is related with the spatial integral including $\delta f$ in Eq.~(\ref{Boltzeq2}).
This integration is also difficult to calculate analytically,
but $G_0(R)$ and $B(R)$ is monotonically decreasing function of $R=\sqrt{R_x^2+R_y^2}$,
and thus
\begin{equation}
\label{}
\hat{\mathcal{I}}(\bm{q}):=\frac{1}{\mathcal{I}}\int\dd^2\bm{R}G_0(R)B(R)e^{i\bm{q}\cdot\bm{R}}=\frac{1}{\mathcal{I}}\int_{-\infty}^\infty\dd R_x\int_{-\infty}^\infty\dd R_yG_0(R)B(R)\cos(q_xR_x)\cos(q_yR_y)
\end{equation}
takes $-1<\hat{\mathcal{I}}(\bm{q})\leq1$.
When $\bm{q}=0$, $\hat{\mathcal{I}}(\bm{q})$ takes the maximum value $\hat{\mathcal{I}}(0)=1$, and when $\bm{q}\rightarrow\infty$, $\hat{\mathcal{I}}(\bm{q})\rightarrow0$.
In the point particle limit $R_D=0$,
\begin{equation}
\label{}
\hat{\mathcal{I}}(\bm{q})=e^{-\frac{q^2R_0^2}{2}}.
\end{equation}

%In addition, this estimated value is on the order of $\mathcal{O}(10^{-5})$ as the maximum value
%for the parameters which we use in this model,
%and therefore the accurate value of the second integral can be very small compared with
%the minimum value of the first integral $\sim\mathcal{O}(10^{-1})$.
%Thus, we can ignore the second integral and obtain
%\begin{equation}
%\label{}
%\mathcal{I}\overline{f_0}\simeq\frac{1-\exp(-\frac{\rho_0R_D^2}{2}\leri{\Phi+\frac{1}{\rho_0R_0^2}})}{\Phi+\frac{1}{\rho_0R_0^2}}.
%\end{equation}

%%%%%%%%%%%%%%%%%%%%%%%%%%%%%%%%%%%%%%%%%%%%%%%%%%%%%%%%%%%%%%
\subsection{Fourier components of the Boltzmann equation}
Except for the term including $\vartheta(\theta_1,\theta_2)$,
the Fourier components ($k,\bm{q}$) of each term of Eq.~(\ref{Boltzeq2}) then become
\begin{equation}
\label{t dv}
\pdv{\delta f(\theta,t)}{t}\rightarrow-\Lambda_k(\bm{q})\delta f_k(\bm{q},t)
\end{equation}
\begin{equation}
\label{nabla dv}
-v_0\bm{e}(\theta)\cdot\grad\delta f(\bm{r},\theta,t)\rightarrow-\frac{v_0}{2}\left\{(q_y+iq_x)\delta f_{k-1}(\bm{q},t)-(q_y-iq_x)\delta f_{k+1}(\bm{q},t)\right\}
\end{equation}
\begin{equation}
\label{s noise}
-s\delta f(\bm{r},\theta,t)+s\int_{-\pi}^\pi\dd\theta'p(\theta-\theta')\delta f(\bm{r},\theta',t)\rightarrow s(2\pi p_k-1)\delta f_k(\bm{q},t),
\end{equation}
\begin{eqnarray}
\label{Iout}
&&-c\mathcal{I}\overline{f_0}\int_{-\pi}^\pi\dd\theta'K(\theta'-\theta)\leri{\delta f(\bm{r},\theta,t)+\frac{1}{\mathcal{I}}\int\dd^2\bm{r}'G_0(\abs{\bm{r}'-\bm{r}})B(\abs{\bm{r}'-\bm{r}})\delta f(\bm{r}',\theta',t)}\\\nonumber
&\rightarrow&-2\pi c\mathcal{I}\overline{f_0}\leri{K_0+\hat{\mathcal{I}}(\bm{q})K_k}\delta f_k(\bm{q},t).
\end{eqnarray}

The term including $\vartheta(\theta_1,\theta_2)$
\begin{equation}
\label{vartheta term}
c\mathcal{I}\overline{f_0}\int_{-\pi}^\pi\dd\theta_1\int_{-\pi}^\pi\dd\theta_2K(\theta_2-\theta_1)\hat{p}(\theta-\vartheta(\theta_1,\theta_2))\leri{\delta f(\bm{r},\theta_2,t)+\frac{1}{\mathcal{I}}\int\dd^2\bm{r}'G_0(\abs{\bm{r}'-\bm{r}})B(\abs{\bm{r}'-\bm{r}})\delta f(\bm{r}',\theta_2,t)},
\end{equation}
is calculated as follows using the same techniques for Eq.~(\ref{Ndf 2}).
The Fourier components of $\bm{q}$ are
\begin{eqnarray}
\label{}
&&2c\mathcal{I}\overline{f_0}\leri{1+\hat{\mathcal{I}}(\bm{q})}\sum_k\sum_{k'}\sum_{k''}K_k\hat{p}_{k'}e^{ik'\theta}\delta f_{k''}(\bm{q},t)\int_{-\pi}^{\pi}\dd\Psi e^{i(2k+k'')\Psi}\{\sgn{\cos\Psi}\}^{k'}\int_{-\pi+\abs{\Psi}}^{\pi-\abs{\Psi}}\dd\Theta e^{-i(k'-k'')\Theta}\\\nonumber
&=&4c\mathcal{I}\overline{f_0}\leri{1+\hat{\mathcal{I}}(\bm{q})}\sum_k\sum_{k'}K_k\hat{p}_{k'}e^{ik'\theta}\delta f_{k'}(\bm{q},t)\int_{-\pi}^{\pi}\dd\Psi e^{i(2k+k')\Psi}\{\sgn{\cos\Psi}\}^{k'}(\pi-\abs{\Psi})\\\nonumber
&&-4c\mathcal{I}\overline{f_0}\leri{1+\hat{\mathcal{I}}(\bm{q})}\sum_k\sum_{k'}\sum_{k''\neq k'}K_k\hat{p}_{k'}e^{ik'\theta}\delta f_{k''}(\bm{q},t)(-1)^{k''}\int_{-\pi}^{\pi}\dd\Psi e^{i(2k+k'')\Psi}\{-\sgn{\cos\Psi}\}^{k'}\frac{\sin((k'-k'')\abs{\Psi})}{k'-k''},
\end{eqnarray}
where the second integral is zero for the same reason as Eq.~(\ref{zero sin}).
We consider only the first integral.
\begin{itemize}
\item
In the case of $k'=-2k$, the first integral is
\begin{eqnarray}
\label{}
&&4c\mathcal{I}\overline{f_0}\leri{1+\hat{\mathcal{I}}(\bm{q})}\sum_kK_k\hat{p}_{-2k}e^{-2ik\theta}\delta f_{-2k}(\bm{q},t)\int_{-\pi}^{\pi}\dd\Psi (\pi-\abs{\Psi})\\\nonumber
&=&4\pi^2c\mathcal{I}\overline{f_0}\leri{1+\hat{\mathcal{I}}(\bm{q})}\sum_kK_k\hat{p}_{-2k}e^{-2ik\theta}\delta f_{-2k}(\bm{q},t).
\end{eqnarray}
\item
In the case of $k'\neq-2k$, the first integral is
\begin{eqnarray}
\label{}
&&4c\mathcal{I}\overline{f_0}\leri{1+\hat{\mathcal{I}}(\bm{q})}\sum_k\sum_{k'\neq-2k}K_k\hat{p}_{k'}e^{ik'\theta}\delta f_{k'}(\bm{q},t)\int_{-\pi}^{\pi}\dd\Psi e^{i(2k+k')\Psi}\{\sgn{\cos\Psi}\}^{k'}(\pi-\abs{\Psi})\\\nonumber
&=&4c\mathcal{I}\overline{f_0}\leri{1+\hat{\mathcal{I}}(\bm{q})}\sum_k\sum_{k'\neq-2k}K_k\hat{p}_{k'}e^{ik'\theta}\delta f_{k'}(\bm{q},t)\times2(-1)^k\left\{(1-(-1)^{k'})\leri{\frac{\pi}{2}\frac{\sin(k'\frac{\pi}{2})}{2k+k'}-\frac{\cos(k'\frac{\pi}{2})}{(2k+k')^2}}\right\}\\\nonumber
&=&4c\mathcal{I}\overline{f_0}\leri{1+\hat{\mathcal{I}}(\bm{q})}\sum_k\sum_{k'\neq-2k}K_k\hat{p}_{k'}e^{ik'\theta}\delta f_{k'}(\bm{q},t)\times2\pi(-1)^k\frac{(-1)^{\frac{k'-1}{2}}}{2k+k'}\delta_{k',\mathrm{odd}}\\\nonumber
&=&8\pi c\mathcal{I}\overline{f_0}\leri{1+\hat{\mathcal{I}}(\bm{q})}\sum_k\sum_{k'}K_k\hat{p}_{k'}e^{ik'\theta}\delta f_{k'}(\bm{q},t)\frac{(-1)^k(-1)^{\frac{k'-1}{2}}}{2k+k'}\delta_{k',\mathrm{odd}}.
\end{eqnarray}
\end{itemize}
Thus, the Fourier components ($k,\bm{q}$) of Eq.~(\ref{vartheta term}) is
\begin{equation}
\label{Iin}
\rightarrow2\pi c\mathcal{I}\overline{f_0}\leri{1+\hat{\mathcal{I}}(\bm{q})}\sum_{k'}K_{k'}(2\pi\hat{p}_{k})\delta f_{k}(\bm{q},t)\leri{\delta_{k,-2k'}+\frac{2}{\pi}\frac{(-1)^{k'}(-1)^{\frac{k-1}{2}}}{2k'+k}\delta_{k,\mathrm{odd}}}.
\end{equation}

%%%%%%%%%%%%%%%%%%%%%%%%%%%%%%%%%%%%%%%%%%%%%%%%%%%%%%%%%%%%%%
\subsection{Calculation of the damping rate}
From Eqs.~(\ref{t dv},\ref{nabla dv},\ref{s noise},\ref{Iout},\ref{Iin}),
we obtain the real part of the damping rate
\begin{equation}
\label{Lambdak}
\Re\Lambda_k(\bm{q})=s(1-2\pi p_{k})+2\pi c\mathcal{I}\overline{f_0}\left\{(K_0+\hat{\mathcal{I}}(\bm{q})K_{k})-2\pi\hat{p}_{k}\leri{1+\hat{\mathcal{I}}(\bm{q})}\sum_{k'}K_{k'}\leri{\delta_{k,-2k'}+\frac{2}{\pi}\frac{(-1)^{k'}(-1)^{\frac{k-1}{2}}}{2k'+k}\delta_{k,\mathrm{odd}}}\right\},
\end{equation}
where we set $v_0=0$ (without the advection term).
When $\Re\Lambda_k(\bm{q})<0$, the uniform distribution becomes unstable.
Note that our aim is the derivation of the transition point,
the cross term of $k\pm1$ modes does not affect to the transition point itself.
In fact, we confirm that this notion is successful,
when we will see later the reproduction of the transition point of the local collision model
as the singular case of our model (see Sect. V).
Essentially, the advection term including the spatial derivative
is not relevant to the transition point. (We provided the reason in the main text.)

Here, 
we assume that $p(\theta)$ (and $\hat{p}(\theta)$) are probability distributions
that satisfies $0<2\pi p_k\leq1$.
For example,
the von Mises distribution
$p(\theta)=e^{\kappa\cos\theta}/(2\pi I_0(\kappa))$ takes $2\pi p_k=I_{\abs{k}}(\kappa)/I_0(\kappa)$
where $I_{n=0,1,\cdots}(\kappa)$ is the modified Bessel function of the first kind
and the sharpness parameter $\kappa>0$,
and the wrapped Gaussian distribution
$p(\theta)=\sum\limits_{n=-\infty}^\infty\frac{1}{\sqrt{2\pi}\sigma}\exp(-\frac{(\theta+2\pi n)^2}{2\sigma^2})$
takes $2\pi p_k=e^{-\sigma^2k^2/2}$.
When we adopt this probability distribution,
the uniform distribution is stable when there is no interaction ($c=0$).

In addition, we find that the uniform distribution is also stable
when the orientational interaction does not depend on the relative angle ($K(\psi)=K_0>0$ and $K_{k\neq0}=0$).
Therefore, $K(\psi)$ must have the deviation from a constant to arise the order.
For $K(\psi)=\sqrt{2}\abs{\sin\psi}$, $K_k$ is 
\begin{equation}
\label{Kk}
K_k=\frac{2\sqrt{2}}{\pi}\frac{\delta_{k,\mathrm{even}}}{1-k^2},
\end{equation}
and then the related term is
\begin{eqnarray}
\label{}
&&2\pi c\mathcal{I}\overline{f_0}\left\{(K_0+\hat{\mathcal{I}}(\bm{q})K_{k})-2\pi\hat{p}_{k}\leri{1+\hat{\mathcal{I}}(\bm{q})}\sum_{k'}K_{k'}\leri{\delta_{k,-2k'}+\frac{2}{\pi}\frac{(-1)^{k'}(-1)^{\frac{k-1}{2}}}{2k'+k}\delta_{k,\mathrm{odd}}}\right\}\\\nonumber
&=&4\sqrt{2}c\mathcal{I}\overline{f_0}\left\{\leri{1+\hat{\mathcal{I}}(\bm{q})\frac{\delta_{k,\mathrm{even}}}{1-k^2}}-2\pi\hat{p}_{k}\leri{1+\hat{\mathcal{I}}(\bm{q})}\sum_{k'}\frac{\delta_{k',\mathrm{even}}}{1-k'^2}\leri{\delta_{k,-2k'}+\frac{2}{\pi}\frac{(-1)^{k'}(-1)^{\frac{k-1}{2}}}{2k'+k}\delta_{k,\mathrm{odd}}}\right\}\\\nonumber
&=&4\sqrt{2}c\mathcal{I}\overline{f_0}\left\{1+\hat{\mathcal{I}}(\bm{q})\frac{\delta_{k,\mathrm{even}}}{1-k^2}-2\pi\hat{p}_{k}\leri{1+\hat{\mathcal{I}}(\bm{q})}\leri{\frac{\delta_{\frac{k}{2},\mathrm{even}}}{1-\leri{\frac{k}{2}}^2}+\frac{2\delta_{k,\mathrm{odd}}}{4-k^2}}\right\},
\end{eqnarray}
where 
\begin{eqnarray}
\label{sum tec}
&&\sum_{k'=\mathrm{even}}\frac{1}{(1-k'^2)(2k'+k)}\\\nonumber
&=&\sum_{n=-\infty}^\infty\leri{\frac{A_1}{4}\frac{1}{n+\frac{k}{4}}-\frac{A_2}{2}\frac{1}{n+\frac{1}{2}}-\frac{A_3}{2}\frac{1}{n-\frac{1}{2}}},~A_1=\frac{4}{4-k^2},~A_2=\frac{2+k}{2(4-k^2)},~A_3=\frac{2-k}{2(4-k^2)}\\\nonumber
&=&\pi\leri{\frac{A_1}{4}\cot(\pi\frac{k}{4})-\frac{A_2}{2}\cot(\frac{\pi}{2})-\frac{A_3}{2}\cot(-\frac{\pi}{2})}\\\nonumber
&=&\pi\frac{A_1}{4}\cot(\pi\frac{k}{4})=\frac{\pi}{4-k^2}(-1)^{\frac{k-1}{2}}.
\end{eqnarray}

Finally, we get
\begin{equation}
\label{}
\Re\Lambda_k(\bm{q})=s\leri{1-2\pi p_k}+4\sqrt{2}c\mathcal{I}\overline{f_0}W_k(\bm{q}),~W_k(\bm{q}):=1+\hat{\mathcal{I}}(\bm{q})\frac{\delta_{k,\mathrm{even}}}{1-k^2}-2\pi\hat{p}_{k}\leri{1+\hat{\mathcal{I}}(\bm{q})}\leri{\frac{\delta_{\frac{k}{2},\mathrm{even}}}{1-\leri{\frac{k}{2}}^2}+\frac{2\delta_{k,\mathrm{odd}}}{4-k^2}}.
\end{equation}
This equation does not depend on the sign of $k$.
We remember $-1<\hat{\mathcal{I}}(\bm{q})\leq1$ and $0<2\pi\hat{p}_k\leq1$,
and then
%and then we investigate for $k>0$.
%Note that we can confirm that $\Lambda_0$ becomes zero,
%because $1-\frac{I_0(\kappa)}{I_0(\kappa)}=0$ and $W_0=0$.
\begin{itemize}

\item
In the case of $k=1$, $W_1(\bm{q})=1-\frac{2}{3}\times(2\pi\hat{p}_1)\leri{1+\hat{\mathcal{I}}(\bm{q})}$.

\item
In the case of  $k=$odd and $k>1$ ($k=3,5,7,\cdots$), $W_k(\bm{q})=1+\frac{2}{k^2-4}(2\pi\hat{p}_1)\leri{1+\hat{\mathcal{I}}(\bm{q})}>0$.

\item
In the case of  $k=$even and $\frac{k}{2}\neq$even ($k=2,6,10,\cdots$), $W_k(\bm{q})=1-\frac{1}{k^2-1}\hat{\mathcal{I}}(\bm{q})>0$.

\item
In the case of  $k=$even and $\frac{k}{2}=$even ($k=4,8,12,\cdots$), $W_k(\bm{q})=1-\frac{1}{k^2-1}\hat{\mathcal{I}}(\bm{q})+\frac{2}{\leri{\frac{k}{2}}^2-1}(2\pi\hat{p}_1)\leri{1+\hat{\mathcal{I}}(\bm{q})}>0$.
\end{itemize}
Therefore, $\Re\Lambda_k(\bm{q})$ can be negative only if $k=1$, and
\begin{equation}
\label{Lamb1}
\Re\Lambda_1(\bm{q})=s\leri{1-2\pi p_1}-4\sqrt{2}c\mathcal{I}\overline{f_0}\left\{\frac{2}{3}\times(2\pi\hat{p}_1)\leri{1+\hat{\mathcal{I}}(\bm{q})}-1\right\}.
\end{equation}
$\hat{\mathcal{I}}(\bm{q})=1$ has a maximum at $\bm{q}=\bm{0}$,
and then $\Re\Lambda_1(\bm{q})$ can be negative in the case of $2\pi\hat{p}_1>3/4$,
e.g. the sharpness parameter of the von Mises distribution is sufficiently large
and the standard deviation parameter of the wrapped Gaussian distribution is sufficiently small.
In this case, $\Re\Lambda_1(\bm{q}=\bm{0})$ can be negative firstly as $c$ increase.

We obtain the transition point by setting $\Re\Lambda_1(\bm{q}=\bm{0})=0$, as
\begin{equation}
\label{ctr sup}
c_{\mathrm{tr}}=\frac{s}{4\sqrt{2}\mathcal{I}\overline{f_0}}\frac{1-2\pi p_1}{\frac{4}{3}\times(2\pi\hat{p}_1)-1}:=\frac{c_0}{\mathcal{I}\overline{f_0}}.
\end{equation}
Furthermore, due to $\hat{\mathcal{I}}(\bm{q})<1$ for $\bm{q}\neq0$,
$\Re\Lambda_1(\bm{q}\neq\bm{0})>0$ at the transition point,
and therefore the any spatial deviation from the uniform distribution attenuates
at the transition point.

%%%%%%%%%%%%%%%%%%%%%%%%%%%%%%%%%%%%%%%%%%%%%%%%%%%%%%%%%%%%%%
\subsection{Limiting cases of the parameters}
Here we consider the behavior of the transition point in the limit cases of parameters $\rho_0,~R_D,~R_0$.

First, when $\rho_0\rightarrow0$, $\mathcal{I}\overline{f_0}$ takes the asymptotic form $\mathcal{I}\overline{f_0}\rightarrow\rho_0R_0^2$
which is consistent with the point particle limit (Eq.~(\ref{point If})) due to reduced occlusion by dilution,
and then $c_{\mathrm{tr}}$ becomes infinity as $c_{\mathrm{tr}}\rightarrow c_0/(\rho_0R_0^2)$.
On the other hand, outside the application of the theory
but we consider $\rho_0\rightarrow\infty$ as a thought experiment,
and $\mathcal{I}\overline{f_0}\rightarrow1/\Phi$,
and thus $c_{\mathrm{tr}}$ has a non-zero value $c_{\mathrm{tr}}\rightarrow c_0\Phi>0$.

Next, we consider the point particle limit $R_D=0$ ($G=1$).
In this case, $\mathcal{I}\overline{f_0}=\rho_0R_0^2$ (see Eq.~(\ref{point If})), and then we obtain the transition point
$c_{\mathrm{tr}}=c_0/(\rho_0R_0^2)$.
When $\rho_0\rightarrow\infty$, $c_{\mathrm{tr}}=0$ for the non-occlusion case.
Therefore, we can confirm that the effect of occlusion ($R_D\neq0$) brings non-zero values of
$c_{\mathrm{tr}}\rightarrow c_0\Phi$.
Moreover, for $R_D=0$, we can directly calculate $\Re\Lambda_1(\bm{q})$ at the transition point.
From Eq.~(\ref{Lamb1}) and Eq.~(\ref{ctr sup}), we obtain
\begin{eqnarray}
\label{Lamb1q}
\Re\Lambda_1(\bm{q})&=&s\leri{1-2\pi p_1}-4\sqrt{2}c_{\mathrm{tr}}\mathcal{I}\overline{f_0}\left\{\frac{2}{3}\times(2\pi\hat{p}_1)\leri{1+e^{-\frac{q^2R_0^2}{2}}}-1\right\}\\\nonumber
&=&2s\frac{1-2\pi p_1}{4\times(2\pi\hat{p}_1)-3}\times(2\pi\hat{p}_1)\leri{1-e^{-\frac{q^2R_0^2}{2}}}>0.
\end{eqnarray}
Therefore, the any spatial deviation from the uniform distribution attenuates
at the phase transition point due to the any slightly non-locality $R_0>0$.

Finally, in the case $R_0\rightarrow\infty$ ($B=1$),
the integrals in $\mathcal{I}\overline{f_0}$ become
\begin{equation}
\label{}
\rho_0\int_0^{R_D}\dd RRe^{-\rho_0\leri{\Phi+\frac{1}{\rho_0R_0^2}}\frac{R^2}{2}}\rightarrow\frac{1}{\Phi}\leri{1-e^{-\frac{\rho_0\Phi R_D^2}{2}}},
\end{equation}
\begin{eqnarray}
\label{}
&&\rho_0\int_{R_D}^\infty\dd RRe^{-\frac{R^2}{2R_0^2}}e^{-\rho_0\Phi R_D^2\left\{-\frac{3}{2}+2\frac{R}{R_D}+\ln(\frac{R_D}{R})\right\}}\\\nonumber
&\rightarrow&\rho_0R_D^2\int_1^\infty\dd \xi\xi\leri{e^{\frac{3}{2}}\xi e^{-2\xi}}^{\rho_0\Phi R_D^2},~\xi=\frac{R}{R_D}\\\nonumber
&=&\rho_0R_D^2e^{\frac{3}{2}\rho_0\Phi R_D^2}E_{-1-\rho_0\Phi R_D^2}(2\rho_0\Phi R_D^2),
\end{eqnarray}
where $E_{a}(z)=\int_1^\infty\dd\xi e^{-z\xi}/\xi^a$ is the generalized exponential integral
and then we can obtain analytically
\begin{equation}
\label{}
\mathcal{I}\overline{f_0}\rightarrow\frac{1}{\Phi}\leri{1-e^{-\frac{\rho_0\Phi R_D^2}{2}}}+\rho_0R_D^2e^{\frac{3}{2}\rho_0\Phi R_D^2}E_{-1-\rho_0\Phi R_D^2}(2\rho_0\Phi R_D^2).
\end{equation}
Thus $c_{\mathrm{tr}}$ takes a non-zero value
even though an agent can perceive a neighbor at infinite relative distance,
due to the visual screening effect.
In fact, at the point particle limit $R_D\rightarrow0$, the generalized exponential integral
has the asymptotic form $E_{-1-\rho_0\Phi R_D^2}(2\rho_0\Phi R_D^2)\rightarrow1/(2\rho_0\Phi R_D^2)^2$,
and then $\mathcal{I}\overline{f_0}\rightarrow\infty$ and $c_{\mathrm{tr}}\rightarrow0$.

\end{widetext}

%%%%%%%%%%%%%%%%%%%%%%%%%%%%%%%%%%%%%%%%%%%%%%%%%%%%%%%%%%%


\begin{thebibliography}{99}

\bibitem{Vicsek2012}
T. Vicsek and A. Zafeiris,
Phys. Rep.
\textbf{517},
71
(2012).

\bibitem{Sridhar2021}
V. H. Sridhar, L. Li, D. Gorbonos, M. Nagy, B. R. Schell, T. Sorochkin, N. S. Gov, and I. D. Couzin,
Proc. Natl. Acad. Sci. U.S.A.
\textbf{118},
e2102157118
(2021).

\bibitem{Peshkin2013}
A. Strandburg-Peshkin, C. R. Twomey, N. W. F. Bode, A. B. Kao, Y. Katz, C. C. Ioannou,
S. B. Rosenthal, C. J. Torney, H. S. Wu, S. A. Levin, and I. D. Couzin,
Curr. Biol.
\textbf{23}, 
R709 
(2013).

\bibitem{Pearce2014}
D. J. G. Pearce, A. M. Miller, G. Rowlands, and M. S. Turner,
Proc. Natl. Acad. Sci. U.S.A.
\textbf{111},
10423
(2014).

\bibitem{Moussaid2011}
M. Moussa\"{i}d, D. Helbing, and G. Theraulaz,
Proc. Natl. Acad. Sci. U.S.A.
\textbf{108},
6884
(2011).

\bibitem{Herbert2011}
J. E. Herbert-Read, A. Perna, R. P. Mann, T. M. Schaerf, D. J. T. Sumpter, and A. J. W. Ward,
Proc. Natl. Acad. Sci. U.S.A.
\textbf{108}, 
18726 
(2011).

\bibitem{Dukas1998}
R. Dukas,
Constraints on information processing and their effects on behavior, in
\textit{Cognitive Ecology: The Evolutionary Ecology of Information Processing
and decision making}
(University of Chicago Press, Chicago, 1998).

\bibitem{Jhawar2020}
J. Jhawar, R. G. Morris, U. R. Amith-Kumar, M. D. Raj, T. Rogers, H. Rajendran, and V. Guttal,
Nat. Phys.
\textbf{16},
488
(2020).

\bibitem{Vicsek1995}
T. Vicsek, A. Czir\'{o}k, E. Ben-Jacob, I. Cohen, and O. Shochet, 
Phys. Rev. Lett. 
\textbf{75}, 
1226 
(1995).

\bibitem{Couzin2002}
I.D. Couzin, J. Krause, R. James, G.D. Ruxton, and N.R. Franks, 
J. theor. Biol. 
\textbf{218}, 
1 
(2002).

\bibitem{chate2008collective}
H. Chat\'{e}, F. Ginelli, G. Gr\'{e}goire, and F. Raynaud,
Phys. Rev. E 
\textbf{77}, 
046113 
(2008).

%\bibitem{Ginelli2010}
%F. Ginelli, F. Peruani, M. B\"{a}r, and H. Chat\'{e}, 
%Phys. Rev. Lett. 
%\textbf{104}, 
%184502 
%(2010).

\bibitem{Lemasson2009}
B. H. Lemasson, J. J. Anderson, and R. A. Goodwin,
J. Theor. Biol.
\textbf{261},
501
(2009).

\bibitem{Kunz2012}
H. Kunz and C. K. Hemelrijk,
Appl. Anim. Behav. Sci.
\textbf{138},
142
(2012).

\bibitem{Bastien2020}
R. Bastien and P. A. Romanczuk,
Sci. Adv.
\textbf{6},
eaay0792
(2020).

\bibitem{Castro2024}
D. Castro, F. Ruffier, and C. Eloy,
Phys. Rev. Research
\textbf{6},
023016
(2024).

\bibitem{Collignon2016}
B. Collignon, A. S\'{e}guret, and J. Halloy,
R. Soc. open sci.
\textbf{3},
150473
(2016).

\bibitem{Gorbonos2024}
D. Gorbonos, N. S. Gov, and I. D. Couzin,
PRX Life
\textbf{2},
013008
(2024).

\bibitem{Ito2024}
S. Ito and N. Uchida,
PNAS Nexus
\textbf{3},
pgae264
(2024).

\bibitem{Toner1995}
J. Toner and Y. Tu,
Phys. Rev. Lett. 
\textbf{75}, 
4326 
(1995).

\bibitem{Toner1998}
J. Toner and Y. Tu,
Phys. Rev. E 
\textbf{58}, 
4828 
(1998).

\bibitem{Bertin2009}
E. Bertin, M. Droz, and G. Gr\'{e}goire,
J. Phys. A: Math. Theor.
\textbf{42}, 
445001 
(2009).

\bibitem{Peshkov2012}
A. Peshkov, S. Ngo, E. Bertin, H. Chat\'{e}, and F. Ginelli,
Phys. Rev. Lett.
\textbf{109}, 
098101 
(2012).

\bibitem{Bertin2015}
E. Bertin, A. Baskaran, H. Chat\'{e}, and M. C. Marchetti,
Phys. Rev. E 
\textbf{92}, 
042141 
(2015).

\bibitem{Patelli2019}
A. Patelli, I. Djafer-Cherif, I. S. Aranson, E. Bertin, and H. Chat\'{e},
Phys. Rev. Lett. 
\textbf{123}, 
258001 
(2019).

\bibitem{Thuroff2014}
F. Th\"{u}roff, C. A. Weber, and E. Frey,
Phys. Rev. X 
\textbf{4}, 
041030 
(2014).

\bibitem{Pita2015}
D. Pita, B. A. Moore, L. P. Tyrrell, and E. Fern\'{a}ndez-Juricic,
PeerJ
\textbf{3},
e1113
(2015).

\bibitem{Calovi2018}
D. S. Calovi, A. Litchinko, V. Lecheval, U. Lopez, A. P. Escudero, H. Chat\'e, C. Sire, and G. Theraulaz, 
PLOS Comput. Biol. 
\textbf{14}, 
e1005933 
(2018).

\bibitem{SUPprob}
We adopt the von Mises distribution to $p(\theta),~\hat{p}(\theta)$,
but our results do not change qualitatively
in the case of the usual probability distribution satisfying
$0<p_k,~\hat{p}_k\leq1/(2\pi)$.
See Appendix 5.

\bibitem{SUPocc} 
See Section I. of Supplementary Material  
for derivation of $\mathcal{D}(R)$.

\bibitem{SUPexG}
See Fig.~S2
for plots of the occlusion factor $G$ for uniform density.


\bibitem{SUPintrinsic} 
See Section III. of Supplemental Material,
where we also confirmed that an alternative form of $K(\psi)$
does not qualitatively change the results.



\bibitem{Bainbridge1958}
R. Bainbridge,
J. Exp. Biol.
\textbf{35},
109
(1958).

\bibitem{Akanyeti2017}
O. Akanyeti, J. Putney, Y. R. Yanagitsuru, G. V. Lauder, W. J. Stewart, and J. C. Liao,
Proc. Natl. Acad. Sci. U. S. A.
\textbf{114},
13828
(2017).

\bibitem{SUPparameter} 
See Section II of Supplemental Material 
for the details of parameter choice.


\bibitem{SUPscaling}
See Section IV of Supplemental Material for derivation of the scaling law,
and Eq.~(S22) for the expression of $P_0(c_{\mathrm{tr}})$.

\bibitem{SUPnonlocal}
See Section V of Supplemental Material 
for the behavior of the model in the local limit $R_0\to 0$.

\bibitem{Martin2021}
D. Martin, H. Chat\'{e}, C. Nardini, A. Solon, J. Tailleur, and F. van Wijland,
Phys. Rev. Lett.
\textbf{126},
148001
(2021);
D. Martin, G. Spera, H. Chat\'{e}, C. Duclut, C. Nardini, J. Tailleur, and F. van Wijland,
J. Stat. Mech.
(2024)
084003.

\bibitem{Rahmani2021}
P. Rahmani, F. Peruani, and P. Romanczuk,
Commun. Phys.
\textbf{4},
206
(2021).

\bibitem{Becco2006}
Ch. Becco, N. Vandewalle, J. Delcourt, P. Poncin,
Physica A
\textbf{367},
487
(2006).

\bibitem{Tunstrom2013}
K. Tunstr{\o}m, Y. Katz, C. C. Ioannou, C. Huepe, and M. J. Lutz, I. D. Couzin,
PLoS Comput. Biol.
\textbf{9},
e1002915
(2013).

\bibitem{Costanzo2018}
A. Costanzo and C. K. Hemelrijk,
J. Phys. D: Appl. Phys.
\textbf{51}, 
134004 
(2018).

\bibitem{Gautrais2009}
J. Gautrais, C, Jost, M. Soria, A. Campo, S. Motsch, R. Fournier, S. Blanco, and G. Theraulaz,
J. Math. Biol. 
\textbf{58}, 
429
(2009).

\bibitem{Ito2022}
S. Ito and N. Uchida,
J. Phys. Soc. Jpn.
\textbf{91}, 
064806 
(2022).

\bibitem{Katz2011}
Y. Katz, K. Tunstr{\o}m, C. C. Ioannou, C. Huepe, and I. D. Couzin,
Proc. Natl. Acad. Sci. U.S.A.
\textbf{108}, 
18720 
(2011).





\end{thebibliography}
\end{document}